\documentclass[%
 reprint,
 amsmath,amssymb,
 aps,nofootinbib,
]{revtex4-1}
\usepackage{graphicx}% Include figure files
\usepackage{dcolumn}% Align table columns on decimal point
\usepackage{bm}% bold math
\PassOptionsToPackage{linktocpage}{hyperref}
\usepackage[hyperindex,breaklinks]{hyperref}
\usepackage{tabularx}
\usepackage{enumitem}
\usepackage{slashed}
\newcommand{\p}{\partial}
\newcommand{\Lag}{\mathcal{L}}
\usepackage{array}
\usepackage{mathtools}
\DeclareMathOperator{\Tr}{Tr}
\usepackage{etoolbox}
\makeatletter
\patchcmd{\@makecaption}{\@ifdim{\wd\@tempboxa >\hsize}}{\@firstoftwo}{}{}
\makeatother

\begin{document}

\preprint{APS/123-QED}

\title{Small neutrino masses from gravitational $\theta$-term}

\author{Gia Dvali$^{1,2,3}$}
\author{Lena Funcke$^{1,2}$}
 \email{Lena.Funcke@physik.uni-muenchen.de}
\affiliation{%
$^1$ Arnold Sommerfeld Center, Ludwig-Maximilians-Universit\"at, Theresienstra{\ss}e 37, 80333 M\"unchen, Germany
}%
 \affiliation{%
$^2$ Max-Planck-Institut f\"ur Physik, F\"ohringer Ring 6, 80805 M\"unchen, Germany
}%
 \affiliation{%
$^3$ Center for Cosmology and Particle Physics, Department of Physics, New York University, 4 Washington Place, New York, NY 10003, USA
}%

\date{\today}

\begin{abstract}
We present how a neutrino condensate and small neutrino masses emerge from a topological formulation of gravitational anomaly. We first recapitulate how a gravitational $\theta$-term leads to the emergence of a new bound neutrino state analogous to the $\eta'$ meson of QCD. Then we show the consequent formation of a neutrino vacuum condensate, which effectively generates small neutrino masses. Afterwards we outline several phenomenological consequences of our neutrino mass generation model. The cosmological neutrino mass bound vanishes since we predict the neutrinos to be massless until the phase transition in the late Universe, $T\sim {\rm meV}$. Deviations from an equal flavor rate due to enhanced neutrino decays in extraterrestrial neutrino fluxes can be observed in future IceCube data. The current cosmological neutrino background only consists of the lightest neutrinos, which, due to enhanced neutrino-neutrino interactions, either bind up, form a superfluid, or completely annihilate into massless bosons. Strongly coupled relic neutrinos could provide a contribution to cold dark matter in the late Universe, together with the new proposed particles and topological defects, which may have formed during neutrino condensation. These enhanced interactions could also be a source of relic neutrino clustering in our Galaxy, which possibly makes the overdense cosmic neutrino background detectable in the KATRIN experiment. The neutrino condensate provides a mass for the hypothetical $B-L$ gauge boson, leading to a gravity-competing force detectable in short-distance measurements. Gravitational waves detections have the potential to probe our neutrino mass generation mechanism.
\end{abstract}

\pacs{14.60.Pq,13.15.+g,04.60.-m,11.30.Rd}

\maketitle

\section{\label{sec:Intro}Introduction}

Numerous theoretical motivations to consider neutrino condensation have been discussed in the literature. In 1967 Ginzburg and Zharkov \citep{Ginzburg1967} pointed out the possibility of a neutrino superfluid in the Universe, but the origin of the neutrino-neutrino interaction was admitted to be unknown. Other works \citep{Barenboim2008, Barenboim2010} suggested that a strongly coupled right-handed neutrino condensate may be the scalar inflaton field, which drives inflation and gives a large mass to the right-handed neutrino. In publications on emergent gravity, for example, in \citep{Moffat2003,McElrath2008}, the long-wavelength fluctuations of the proposed neutrino condensate were identified with a Goldstone graviton. The possible connection between a neutrino condensate and dark energy has also been pointed out, see, e.g., \citep{Moffat2003,Bhatt2009,Azam2010}. However, the considered attractive neutrino interactions providing the condensate were either not known or based on the proposed existence of additional fields or new physics at high-energy scales. 

In the current paper, neutrino condensation directly emerges from a topological formulation of gravitational anomaly. In order to clarify our argument, we will first outline the analogy to well-known QCD effects. 

In QCD, the topologically nontrivial $\theta$-vacuum spontaneously breaks chiral symmetry. This gives rise to eight pseudo-Goldstone bosons, the mesons. Due to the Adler-Bell-Jackiw (ABJ) anomaly of the isospin singlet axial current \citep{Adler1969,Bell1969}, the ninth meson $\eta'$ is much heavier than its eight companions and cannot be regarded as a pseudo-Goldstone boson. This mass splitting is known as the $\eta'$ puzzle and has two different resolutions: the 't Hooft instanton mechanism \citep{'tHooft1976} and the Witten-Veneziano mechanism \citep{Witten1979,Veneziano1979}. The effect that will serve as an important analogy for our neutrino consideration takes place in case when at least one of the quarks has vanishing bare mass. As it is well known, in this case the vacuum $\theta$-angle of QCD becomes unphysical, since it can be rotated away by an anomalous chiral rotation of the massless quark field. This case is important for us because of its alternative description in a topological language.

As pointed out in \cite{Dvali2005_2} using the power of gauge redundancy, we can understand the elimination of the vacuum $\theta$-angle in terms of a three-form Higgs effect,  in which the $\eta'$ meson is eaten up by the topological QCD Chern-Simons three-form, and the two combine into a single massive pseudoscalar particle. In \cite{Dvali2005} it was shown that this phenomenon can be formulated in a model-independent way, entirely in terms of topology and anomaly, without the need of knowledge of the underlying microscopic structure of the theory: whenever a theory contains a vacuum $\theta$-angle, which can be eliminated by chiral transformation, the mass gap is necessarily generated and there exists a pseudo-Goldstone boson, which is eaten up by a corresponding Chern-Simons three-form. The generality of the phenomenon makes it applicable to other systems, such as gravity coupled to neutrino species.

Indeed, in \cite{Dvali2005_2} it was pointed out that in case of the existence of physical vacuum $\theta$-angles, gravity and QCD have a similar topological structure: the gravitational Chern-Simons three-form enters the Higgs phase provided the theory contains a fermion with zero bare mass, such as the neutrino. As shown in \citep{Dvali2013}, an important consequence of this three-form Higgs effect follows for the neutrinos: they are the analogon to the light quarks in QCD, and consequently the neutrino sector delivers a pseudo-Goldstone boson of broken axial neutrino symmetry, $\eta_{\nu}$, which becomes a longitudinal component of the Chern-Simons gauge three-form and generates a mass gap in the theory. This pseudo-Goldstone boson represents a bound neutrino state triggered by gravitational anomaly, analogous to the $\eta'$ triggered by the ABJ anomaly of QCD. 

In order to elucidate the topological similarities of gravity and QCD, we will recapitulate these ideas in Sec. \ref{sec:MassGap}. Moreover, in that section we will derive that an order parameter, most obvious a neutrino vacuum condensate triggered by nonperturbative gravitational effects, is required in order to deliver the new degree of freedom $\eta_\nu$.

The predicted new bound neutrino state implies a further important theoretical consequence for the neutrino sector, which we will present in Sec. \ref{sec:MassGen}: the same neutrino vacuum condensate that delivers the degree of freedom $\eta_\nu$ also generates the small neutrino masses. We thus establish the origin of the small neutrino mass as the accompanying phenomenon to the  elimination of the gravitational  $\theta$-term by the axial neutrino anomaly.

Numerous ways to explain the small neutrino masses have already been evaluated, for example, through the celebrated see-saw mechanism \citep{Minkowski1977,Gell-Mann1979,Yanagida1980,Mohapatra1980,Schechter1980,Schechter1982} and through radiative corrections (see, e.g., the Zee model \citep{Zee1980} or Witten's right-handed neutrino mass generation in $SO(10)$ \citep{Witten1980}), or in models with large extra dimensions \cite{ArkaniHamed1998,Dienes1998,Dvali1999}. Our mass generation mechanism is especially interesting since it is independent of the Majorana or Dirac nature of the neutrinos and only bases on one single widespread \citep{Kallosh1995} assumption: that the gravitational $\theta$-term is physical in the absence of massless chiral fermions. 

Our model has far-reaching phenomenological consequences, which are elucidated in Sec. \ref{sec:PhenImplications}, such as the invalidity of the cosmological neutrino mass bound, enhanced neutrino-neutrino interactions, and neutrino decays. 

In Sec. \ref{sec:Conclusion} we will summarize the expected experimental signals and discuss the theoretical significance of our idea, including its implications on $CP$ violation by the gravitational vacuum and in gravitational waves.

In the whole paper we will omit irrelevant numerical factors and reintroduce them only if necessary.

\section{\label{sec:MassGap}Bound states from anomalies}

In \citep{Dvali2005} it was shown that in the presence of an anomaly, the topological formulation of any model immediately generates a mass gap in the theory, rendering the corresponding vacuum $\theta$-angle unphysical. In particular, this absolutely generic concept illuminates the origin of the massive $\eta'$ degree of freedom in QCD in model-independent terms. In the formulation of \cite{Dvali2005_2}, the power of gauge redundancy allows to understand this phenomenon as a Higgs effect of the corresponding Chern-Simons three-form. 

It was pointed out in \cite{Dvali2005_2} that the similar effect of a mass gap generation must be exhibited by gravity in the presence of neutrinos with zero bare mass. The authors of \citep{Dvali2013} showed that in the presence of an anomaly, gravity gives rise to a new degree of freedom in the neutrino sector, a pseudo-Goldstone boson of spontaneously broken axial neutrino symmetry. This $\eta_\nu$ particle is analogous to the $\eta'$ meson of QCD. 

We will recapitulate the theoretical foundations of the mass gap generation from anomalies in the cases of QCD and gravity in the following two subsections, respectively.

\subsection{\label{sec:QCDMassGap}Topological mass gap generation in QCD}

In QCD, the nontrivial vacuum spontaneously breaks the chiral quark symmetry,
\begin{equation*}
U(3)_L \times U(3)_R \rightarrow U(3)_{V} =SU(3)_{V} \times U(1)_{V},
\end{equation*}
which in principle should give rise to $8+1$ pseudo-Goldstone bosons. However, in nature we only observe eight light mesons, and the ninth singlet pseudoscalar meson $\eta'$ is much heavier than its companions. This is due to the fact that the corresponding isospin singlet axial $U(1)$ current 
\begin{equation}
j_5^\mu=\bar{q}\gamma^\mu \gamma_5 q
\end{equation}
has an anomalous divergence found by ABJ \citep{Adler1969,Bell1969},
\begin{equation}\label{eq:QCDj}
\partial_\mu j_5^\mu=G\tilde{G}+m_{q}\bar{q} \gamma_5 q,
\end{equation}
where $G$ is the gluon field strength and $\tilde{G}$ is its Hodge dual,
\begin{align}
G^a &\equiv dA^a + \, f^{abc} A^b A^c, \\ 
\tilde{G}^{a\;\alpha\beta} &\equiv \epsilon^{\alpha\beta\mu\nu} G^a_{\mu\nu}.
\end{align}
Here, $G=G^aT^a$, $A$ is the gluon field matrix, $d$ denotes the exterior derivative, $T^a$ are the generators and $f^{abc}$ are the structure constants of the appropriate Lie algebra.

It is well known \citep{Dvali2005} that one can equivalently and more elegantly formulate QCD in terms of topological entities: the Chern-Simons three-form $C$  and the Chern-Pontryagin density $E$,
\begin{align}
C &\equiv AdA - {3 \over 2} AAA,\label{eq:C}\\ 
E &\equiv G\tilde{G}=\mathrm{d}C.\label{eq:E}
\end{align}
The Chern-Simons three-form $C$ \eqref{eq:C} obtains the meaning of a field in QCD and plays a decisive role in the infamous strong \textit{CP} problem. QCD is $\theta$-dependent only if its topological vacuum susceptibility $\langle G\tilde{G},G\tilde{G}\rangle$ does not vanish in the limit of zero momentum,
\begin{align}\label{eq:FFcorr}
\langle G\tilde{G},G\tilde{G}\rangle_{q\to 0}&\equiv\lim\limits_{q \to 0}\int d^4 x\: e^{iqx} \langle T[G\tilde{G}(x)G\tilde{G}(0)]\rangle\nonumber\\
&=\mathrm{const}\neq 0.
\end{align}
Expressed in terms of the field $C$, the nonvanishing correlator \eqref{eq:FFcorr} implies that $C$ has a massless pole for vanishing momentum,
\begin{equation}\label{eq:CCcorr}
\langle C, C\rangle_{q\to 0}=\frac{1}{q^2},
\end{equation}
implying that $C$ is a massless gauge field. This fact combined with the gauge symmetry of $C$ leads to the following effective Lagrangian describing the topological structure of the QCD vacuum, which is simply a gauge theory of the massless three-form $C$ \cite{Dvali2005_2}:
\begin{equation}
\label{effQCD}
\Lag=\frac{1}{\Lambda^4}E^2 +  \theta E + {\rm higher ~order~ terms},
\end{equation}
where the QCD scale $\Lambda$ takes care of the dimensionality. Notice that in this formulation the famous $\theta$-term
\begin{equation}\label{eq:L}
\Delta \Lag=\theta G\tilde{G}
\end{equation}
in the vacuum is nothing but the vacuum expectation value of the electric field, $\theta \langle G\tilde{G}\rangle=\theta\langle E \rangle = \theta \Lambda^4$. 

Inserting Eq. \eqref{eq:E} into Eq. \eqref{eq:L} shows that the $\theta$-term in the QCD Lagrangian is a total derivative. As 't Hooft, Witten, and Veneziano pointed out \citep{'tHooft1976,Witten1979,Veneziano1979}, only nonperturbative effects give this term a physical significance. For a good review, see, for example, \citep{Bass2008}.

If we now introduce massless quarks (or axions \citep{Weinberg1977,Wilczek1978}) into the theory in order to make QCD independent of $\theta$, the topological susceptibility of the QCD vacuum \eqref{eq:FFcorr} vanishes and the electric field $E$ \eqref{eq:E} gets screened. Consequently, the massless pole of the low-energy correlator $\langle C, C\rangle_{q\to 0}$ \eqref{eq:CCcorr} has to be eliminated, which is only possible through introducing a mass gap into the theory,
\begin{equation}\label{eq:CGap}
\langle C, C\rangle_{q\to 0}=\frac{1}{q^2+m^2}.
\end{equation}

By gauge symmetry, the only way to account for this phenomenon in the language of the effective Lagrangian (\ref{effQCD}) is by introducing a massive pseudoscalar degree of freedom, which generates the mass gap. If the chiral symmetry was due to massless quarks, the corresponding massive pseudo-Goldstone boson would be $\eta'$. By symmetries, the lowest-order terms are then uniquely fixed to be
\begin{equation}\label{effectiveQCDeta}
\Lag = \frac{1}{2\Lambda^4} E^2 -\frac{1}{f_{\eta'}}\eta'E + {1 \over 2} \p_\mu \eta'\p^\mu\eta',
\end{equation}
where the decay constant $f_{\eta'}$ of the $\eta'$ meson is given by the QCD scale. 

Solving the equation of motion for $C$,
\begin{equation}\label{eqEeff}
\mathrm{d}\left( E - {\Lambda^4 \over f_{\eta'}} \eta' \right)=0,
\end{equation}
we obtain for the electric field
\begin{equation}\label{solEeff}
E=\Lambda^4\left({\eta'  \over f_{\eta'}}-\theta\right),
\end{equation}
where $\theta$ is an integration constant.

Inserting this in the equation of motion for $\eta'$ we get 
\begin{equation}\label{boxeta}
\Box \eta' +  {{\Lambda^4} \over f_{\eta'} }\left({\eta' \over f_{\eta'}} -\theta \right) = 0. 
\end{equation}
Here we immediately observe two important points:

\begin{enumerate}[label={(\arabic*)}]
\item The fields $\eta'$ and $C$ combine and form a single propagating massive bosonic field of mass $m_{\eta'} = \Lambda^2 / f_{\eta'}$. 
\item The vacuum expectation value of this field is exactly where the electric field $E$ vanishes. 
\end{enumerate}

It is clearly seen that this low-energy perspective of screening the electric field $E$, rendering the $\theta$-term unphysical and solving the strong \textit{CP} problem, matches the high-energy perspective: $\theta$ becomes an unobservable quantity since a change in the effective value of $\theta$ can always be induced by making a chiral rotation of the massless fermion fields. 

Let us emphasize this point again: expressed in the topological three-form language, the solution of the strong \textit{CP} problem equals to the generation of a mass gap for $C$ \citep{Dvali2005_2}. This is exactly what happens when introducing massless quarks or an additional Peccei-Quinn (PQ) symmetry \citep{Peccei1977} into the theory. The emerging massive pseudoscalar degree of freedom, which is either the $\eta'$ in the massless quark case ($m=m_{\eta'}$) or the axion in the PQ solution ($m=m_{a}$), is eaten up by $C$. This means that the pseudoscalar provides a mass for the three-form and is the origin of the massive pole of the correlator $\langle C, C\rangle_{q\to 0}$ \eqref{eq:CGap}.

\subsection{\label{sec:GravityMassGap}Topological mass gap generation from gravitational anomaly}

Let us now consider the extension of Einstein gravity by a unique Chern-Simons term $\theta_G R\tilde{R}$ in the Lagrangian \citep{Jackiw2003}. The general considerations of the previous subsection can then be directly applied to gravity, as accomplished in \citep{Dvali2005_2,Dvali2013}. Analogous to QCD we can formulate gravity in terms of topological quantities: a gravitational Chern-Simons three-form $C_G$ and a gravitational Chern-Pontryagin density $E_G$,
\begin{align}
C_G&\equiv \Gamma \mathrm{d}\Gamma-\frac{3}{2} \Gamma\Gamma\Gamma,\label{eq:gravcs}\\
E_G&\equiv R\tilde{R}=\mathrm{d}C_G.\label{eq:gravfieldstr}
\end{align}
Here, $\Gamma$ is the Christoffel connection, $R$ is the Riemann tensor, and $\tilde{R}$ is its Hodge dual.

Our starting assumption is that in the absence of massless fermions (for example in a theory with pure gravity) the gravitational vacuum angle $\theta_G$ would be physical, i.e., the topological vacuum susceptibility would not vanish, 
\begin{equation}\label{eq:RRcorr}
\langle R \tilde{R}, R\tilde{R}\rangle_{q\to 0}=\mathrm{const}\neq 0.
\end{equation}
The strength of the correlator  $\langle R \tilde{R}, R\tilde{R}\rangle_{q\to 0}$ is given by a scale $\Lambda_G$, which will appear as an effective cutoff scale in a low-energy theory of the gravitational three-form $C_G$. The scale $\Lambda_G$ is unknown and will be treated as a parameter, solely fixed from phenomenological requirements.

In the following we will consider gravity coupled to the lightest known fermions, the neutrinos, in analogy to considering QCD with light quarks. Just like in the case of quarks, there exists a neutrino chiral symmetry, which is anomalous under gravity. If no right-handed neutrinos are introduced, i.e., if neutrinos are purely Majorana particles, this anomalous symmetry coincides with the neutrino lepton number. In case when neutrinos have right-handed partners, they carry the opposite charges under the anomalous symmetry, which is different from the standard lepton number charge assignment. Hence, to stress this difference we will call this symmetry an \textit{axial} neutrino lepton number. 
 
In what follows we will outline how the anomalous axial neutrino symmetry leads to a massive new pseudoscalar degree of freedom. For simplicity, we will first consider a single massless neutrino species and afterwards add a small neutrino mass as a perturbation.

As already mentioned, the axial $U(1)$ neutrino current
\begin{equation}
\label{eq:neutcurr}
j_5^\mu=\bar{\nu}\gamma^\mu \gamma_5\nu
\end{equation}
corresponding to the axial neutrino symmetry 
\begin{equation}
\label{eq:chirsymm}
\nu\to e^{i\gamma_5\chi}\nu
\end{equation}
has an anomalous divergence \citep{Delbourgo1972,Eguchi1976,Deser1980,AlvarezGaume1984}
\begin{equation}
\label{eq:divneutJ}
\p_\mu j^{\mu}_5= R\tilde{R}= E_G.
\end{equation}
Due to this anomaly, the effective Lagrangian of the gravitational three-form field $C_G$ is given by
\begin{equation}
\label{eq:efflagcur}
\Lag=\frac{1}{2\Lambda_G^4}E_G^2+\frac{1}{f_{\nu}^2}E_G\frac{\p_\mu}{\Box}j_5^{\mu}.
\end{equation}
The first term accounts for treating the field $C_G$ in an effective low-energy theory, where terms of higher order in $E_G$ and its derivatives can be neglected. The unique contact interaction between $j_{5}$ and $E_G$ with strength $f_{\nu}\sim \Lambda_G$ is generated by the triangle diagram of the gravitational ABJ anomaly \citep{Dvali2005_2}. 
 
The equation of motion for $C_G$,
\begin{equation}
\left(\Box+\Lambda^2_G\right) E_G=0,
\end{equation}
shows that there are no massless modes in $E_G$. This means that massless neutrinos can screen the gravitational electric field $E_G$ and hence make the gravitational $\theta$-term vanish, completely analogous to the massless quark case in QCD. The corresponding generation of the mass gap for $C_G$, analogous to Eq. \eqref{eq:CGap}, automatically implies that the current \eqref{eq:neutcurr} must be identified with a pseudoscalar degree of freedom. We shall call it $\eta_{\nu}$ in analogy to the $\eta'$ meson of QCD.

It is important to notice that the existence of $\eta_\nu$ is required for generating the mass gap in the presence of gravitational anomaly \citep{Dvali2005_2,Dvali2005,Dvali2013}. The correlator \eqref{eq:RRcorr} has to be screened, which can only be accomplished if $C_G$ eats up a Goldstone-like degree of freedom and gets massive, i.e., the St\"uckelberg field $\eta_\nu$ \textit{has} to arise in order to preserve gauge symmetry. As already mentioned in the previous subsection, this three-form Higgs effect is the low-energy perspective of rendering the gravitational $\theta$-term unphysical. From the high-energy point of view, $\theta_G$ is made unobservable since it can be arbitrarily shifted by a chiral rotation of the massless neutrino field.

By analogy with $\eta'$, $\eta_\nu$ can be regarded as a pseudo-Goldstone boson arising since nonperturbative gravity breaks axial neutrino number symmetry \eqref{eq:chirsymm}. It can be written as the effective low-energy limit of the neutrino bilinear operator
\begin{equation}\label{eq:etanudef}
\eta_{\nu} \rightarrow \frac{1}{\Lambda_G^2}\bar{\nu}\gamma_5\nu \quad
\end{equation}
with the corresponding axial singlet current \eqref{eq:neutcurr}
\begin{equation}\label{eq:currenteta}
\quad j_5^{\mu}\rightarrow \Lambda_G\p^\mu\eta_{\nu}.
\end{equation}
The effect of the mass gap generation is readily accounted by the effective Lagrangian for $\eta_{\nu}$, which is obtained by inserting Eq. (\ref{eq:currenteta}) into Eq. (\ref{eq:efflagcur}),
\begin{equation}\label{eq:effective}
\Lag = \frac{1}{2\Lambda_G^4} E_G^2 -\frac{1}{\Lambda_G}\eta_{\nu} E_G + {1 \over 2} \p_\mu \eta_{\nu}\p^\mu\eta_{\nu},
\end{equation}
analogous to the QCD Lagrangian \eqref{effectiveQCDeta}.

Since the decay constant of the $\eta'$ meson is given by the QCD scale \citep{Dvali2013}, we here analogously identify the decay constant $f_{\nu}$ of $\eta_{\nu}$ with $\Lambda_G$.

The equation of motion for $\eta_{\nu}$, 
\begin{equation}\label{eq:boxeta}
\Box \eta_{\nu} +  \frac{1}{\Lambda_G} E_G = 0, 
\end{equation}
immediately implies that $E_G=R \tilde{R}$ must vanish in any state in which $\eta_{\nu}$ is constant and, in particular, in the vacuum. 

After integrating the equation of motion for $E_G$, 
\begin{equation}\label{eq:eomforCGneut}
\mathrm{d}\left( E_G -\Lambda_G^3\eta_{\nu}\right)=0,
\end{equation}
we obtain for the electric field
\begin{equation}\label{eq:E_G}
E_G=\Lambda_G^3(\eta_{\nu}-\theta_G\Lambda_G),
\end{equation}
where $\theta_G$ is an integration constant. 

Inserting this expression for $E_G$ into Eq. (\ref{eq:boxeta}), we get
\begin{equation}
\Box \eta_{\nu} +  \Lambda_G^2(\eta_{\nu}-\theta_G \Lambda_G) = 0.
\end{equation}
We see that $\eta_{\nu}$ is a massive field with the vacuum expectation value exactly at the point $\eta_{\nu} =  \theta_G \Lambda_G$, which makes $E_G$ zero in the vacuum.

{\renewcommand{\arraystretch}{1.2}
\begin{table*}[t]
  \centering
\begin{tabular*}{\textwidth}{l @{\extracolsep{\fill}} l l}
   \hline
   \hline
   \multicolumn{1}{c}{ \rule{0pt}{3ex} Quantity} & \multicolumn{1}{c}{QCD} & \multicolumn{1}{c}{Gravity} \\
   \hline
Anomalous axial $U(1)$ symmetry & $q\to \exp(i\gamma_5\chi)q$ & $\nu\to \exp(i\gamma_5\chi)\nu$ \\
Anomalous axial $U(1)$ current & $j_5^\mu=\bar{q}\gamma^\mu \gamma_5 q$ & $j_5^\mu=\bar{\nu}\gamma^\mu \gamma_5\nu$ \\
Corresponding anomalous divergence & $\partial_\mu j_5^\mu=G\tilde{G}+m_{q}\bar{q} \gamma_5 q$ & $\p_\mu j_5^\mu= R\tilde{R} + m_\nu\bar{\nu}\gamma_5\nu$\\
Corresponding pseudoscalar & $\eta' \rightarrow \bar{q}\gamma_5q /\Lambda^2$ & $\eta_{\nu} \rightarrow \bar{\nu}\gamma_5\nu /\Lambda_G^2$\\
Chern-Simons three-form & $C \equiv AdA - {3 \over 2} AAA,$ & $C_G\equiv \Gamma \mathrm{d}\Gamma-\frac{3}{2} \Gamma\Gamma\Gamma$\\
Chern-Pontryagin density & $E \equiv G\tilde{G}=\mathrm{d}C$ & $E_G\equiv R\tilde{R}=\mathrm{d}C_G$\\
Topological vacuum susceptibility & $\langle G \tilde{G}\rangle_{q\to 0}=-\theta m_q \langle \bar{q}q\rangle$ & $\langle R \tilde{R}\rangle_{q\to 0}=-\theta_G m_\nu \langle \bar{\nu}\nu\rangle$ \\
    \hline
    \hline
    \end{tabular*}
    \caption{Overview of the analogy between QCD and gravity. For simplicity, only a single fermion flavor is considered.}\label{tab:1}
\end{table*}
}

From the effective Lagrangian \eqref{eq:effective} it is clear that an axial rotation of the neutrinos results into the shift of $\eta_{\nu}$ by a constant. This conforms the self-consistency of the statement that $\eta_{\nu}$ is a pseudo-Goldstone boson of the spontaneously broken axial neutrino symmetry. It is obvious to identify the order parameter of this symmetry breaking with a neutrino condensate, similar to the quark condensate in QCD. The validity of this identification will be further discussed below, but at this point is only secondary. What is important here is that there exists an order parameter, which breaks axial neutrino symmetry, and that the corresponding pseudo-Goldstone boson $\eta_{\nu}$ makes the Chern-Pontryagin density $E_G$ zero. 

The similarity of the gravity and QCD stories continues also in the presence of a small bare neutrino mass, which breaks axial symmetry explicitly. In order to see this, we now introduce a bare neutrino mass $m_{\nu}$ into the picture, which provides a small explicit mass for $\eta_{\nu}$ in the effective low-energy Lagrangian. Analogous to the small explicit $\eta'$ mass given by the $u$, $d$, and $s$ quark masses, the small explicit $\eta_\nu$ mass is not related to the anomaly and vanishes in the chiral limit. 

In the presence of a bare neutrino mass, the axial neutrino number is explicitly broken, and the divergence of the singlet axial current \eqref{eq:neutcurr} obtains a second contribution, just as in the QCD case \eqref{eq:QCDj},
\begin{equation}\label{eq:neutcurrentmass}
\p_\mu j_5^\mu= R\tilde{R} + m_\nu\bar{\nu}\gamma_5\nu.
\end{equation}

Again replacing the current \eqref{eq:neutcurrentmass} in the initial Lagrangian (\ref{eq:efflagcur}) with the pseudo-Goldstone boson \eqref{eq:etanudef}, the effective Lagrangian (\ref{eq:effective}) acquires an additional explicit mass term for $\eta_{\nu}$,
\begin{equation}\label{eq:etamassLag}
\hspace*{-0.1cm}\Lag = \; \frac{1}{2\Lambda_G^4} E_G^2 -\frac{1}{\Lambda_G}\eta_{\nu} E_G + {1 \over 2} \p_\mu \eta_{\nu}\p^\mu\eta_{\nu}- {1 \over 2} m_{\nu} \Lambda_G\eta_{\nu}^2.
\end{equation}
The equation of motion and the corresponding solution for $E_G$ are again given by Eqs. (\ref{eq:eomforCGneut}) and (\ref{eq:E_G}), but the equation for $\eta_{\nu}$ changes to 
\begin{equation}\label{boxeta}
\Box \eta_{\nu} +  \Lambda_G^2(\eta_{\nu}-\theta_G \Lambda_G) + m_{\nu} \Lambda_G\eta_{\nu} = 0, 
\end{equation}
so that the vacuum expectation value of $\eta_{\nu}$ is shifted to 
\begin{equation}
\eta_{\nu} = \frac{\theta_G \Lambda_G}{1 + \frac{m_{\nu}}{\Lambda_G}}.
\end{equation}
Inserting this in Eq. (\ref{eq:E_G}), we get in the leading order in $m_{\nu} / \Lambda_G$ expansion
\begin{equation}
\langle R \tilde{R}\rangle_{q\to 0} = \langle E_G \rangle_{q\to 0} \simeq  - \theta_G m_{\nu} \Lambda_G^3.
\end{equation}
Expressing the order parameter of the spontaneous axial symmetry breaking with the neutrino condensate $\Lambda_G^3  = \langle \bar{\nu}\nu\rangle$, we can write the vacuum expectation value of the gravitational Chern-Pontryagin density in the form
\begin{equation}\label{eq:NuVacuum}
\langle R \tilde{R}\rangle_{q\to 0}= - \theta_G m_\nu \langle \bar{\nu}\nu\rangle.
\end{equation}

With a single basic assumption that in the absence of massless chiral fermions the gravitational $\theta$-term Chern-Pontryagin density would be physical, we uniquely arrive to a story very similar to QCD. Namely, once massless neutrinos couple to gravity, they condense and deliver a pseudo-Goldstone boson $\eta_{\nu}$, which generates a mass gap and screens the Chern-Pontryagin density, rendering the gravitational $\theta$-term unphysical. 

The absence of the condensate in a theory with massless fermions would cause an obvious contradiction. On the one hand, since the $\theta$-term can be rotated away by an axial transformation, the Chern-Pontryagin density must vanish. On the other hand, this requires the generation of a mass gap for the Chern-Simons three-form field, for which the existence of a to-be-eaten-up pseudo-Goldstone boson $\eta_{\nu}$ is necessary. But in the absence of a condensate, which breaks axial symmetry spontaneously, the origin of such a Goldstone boson is impossible to explain.

There is a second argument why the identification of the order parameter with a neutrino vacuum condensate is plausible. The authors of \citep{Shifman1979} computed the QCD vacuum susceptibility to linear order in the $u$ and $d$ quark masses,
\begin{align}\label{eq:Vacuum}
\langle G \tilde{G}\rangle_{q\to 0}&= \theta \langle G \tilde{G}, G\tilde{G}\rangle_{q\to 0}\nonumber\\
&= - \theta  \frac{m_u m_d}{(m_u+m_d)^2} \langle m_u\bar{u}u+m_d\bar{d}d\rangle.
\end{align}
The crucial point in their derivation is that the mass of the $\eta'$ meson does not vanish in the chiral limit, i.e., that it has an additional mass contribution from the anomaly. In \citep{Shifman1979} it is strongly emphasized that the computation of Eq. \eqref{eq:Vacuum} is unrelated to confinement and does not depend on the theory's coupling. If ones recapitulates the computation step by step, it becomes clear that specific QCD characteristics such as the soft pion technique are not required at all for the derivation of Eq. \eqref{eq:Vacuum}. Therefore, the computation and argumentation presented in \citep{Shifman1979} are transferable to gravity. In gravity, the mass of the $\eta_\nu$ also does not vanish in the limit of zero bare neutrino mass. Thus, in the simplified case of only one fermion flavor, we can directly transfer Eq. \eqref{eq:Vacuum} to the gravitional neutrino sector and observe that the resulting expression for the gravitational vacuum susceptibility coincides with the result of our derivation, Eq. \eqref{eq:NuVacuum}. This provides another strong hint that our identification of the order parameter with a neutrino vacuum condensate triggered by nonperturbative gravitational effects is valid.

At this point it is important to notice that the neutrino bound state $\eta_{\nu}$ differs from bound states arising due to universal confining dynamics. The $\eta_{\nu}$ is forced upon us by the Goldstone theorem and the three-form Higgs effect, which is due to the same nonperturbative gravitational dynamics that is responsible for the $\langle R\tilde{R},R\tilde{R}\rangle$ correlator in the massive-fermion theory and for its screening in the massless one. These dynamics are not obliged to produce universal gravitational confinement among other particles.

However, the neutrino condensate possibly requires neutrinos to bind up below the symmetry breaking scale $\Lambda_G$, due to 't Hooft's anomaly matching condition.\footnote{The authors thank N. Wintergerst for this comment.} If we assume neutrinos to be in bound states, we can distinguish two options. One is the binding of only low-energy neutrinos below $\Lambda_G$ energies. In the next section we will identify $\Lambda_G$ with the neutrino mass scale; thus, it would be consistent with current observations that no free neutrinos exist below this low-energy scale. The second option is to consider a classic picture in which neutrinos are connected by nonperturbative flux tubes of tension of order (meV)$^2$. Such tubes would be hard to rule out based on existing observations, for example, neutrinos of MeV energy can stretch the flux tube to a huge macroscopic length $L \sim 10^{7}~{\rm cm}$. 

So far we have only considered one single neutrino species. Incorporating three neutrino flavors is straightforwardly possible and extends the result \eqref{eq:NuVacuum}. In this case we obtain additional pseudo-Goldstone bosons $\phi_{k}$ if we assume the neutrinos to have hard masses smaller than the symmetry breaking scale $\Lambda_G$. If neutrinos have no hard but only effective masses, i.e., if the only source of all the observed neutrino masses is the spontaneous breaking of axial neutrino symmetry by the neutrino condensate, as proposed in the next section, then the $\phi_k$'s become massless Goldstones\footnote{Even though it might seem counterintuitive to write a massless Goldstone particle as a bilinear of effectively massive constituents, one should be reminded that Goldstone bosons cannot simply be regarded as bound states of elementary particles. This distinguishes our new degrees of freedom from, e.g., WIMP-onium states \citep{Shepherd2009}. In case of neutrinos with no Yukawa couplings the chiral symmetry is not explicitly broken. We consequently have to obtain eight massless and one massive new degree of freedom, no matter how counterintuitive. (For analogous theoretical considerations, see, e.g., the experimentally ruled-out top condensate model \citep{Bardeen1989} based on \citep{Nambu1961}.) In this discussion we ignore small corrections to the masses of some of the Goldstones generated by weak-scale effects.}. The number of additional Goldstone modes depends on the pattern of symmetry breaking. The condensate of Dirac neutrino flavors transforms as bifundamental under $U(3)_L \times U(3)_R$ symmetry and can potentially break it down to a diagonal $U(1)_1\times U(1)_2 \times U(1)_3$ subgroup of individual neutrino number symmetries for each flavor. This would result into $14$ massless Goldstones $\phi_{k}$ and one massive pseudo-Goldstone $\eta_\nu$. The off-diagonal Goldstones can induce neutrino-flavor-changing transitions. In the following we will denote the massless Goldstones together with the massive $\eta_\nu$ as $\phi\equiv\{\phi_{k},\eta_\nu\}$. 

Here at the end of the section we want to face the question why the analogy between QCD and gravity works so well, even though these two theories seem to be completely different on the first sight. How can we at all perform computations in quantum gravity and consider the analogy to QCD, even though quantum gravity still requires a consistent UV completion? The answer is simple: anomalies are only sensitive to the massless sector of a theory and hence are insensitive to its energy scale \citep{AlvarezGaume1984,'tHooft1980}, so we do not need to understand the UV regime of quantum gravity and can safely work in the well-understood effective low-energy regime. 

Lastly, we point out that Table \ref{tab:1} provides a concise overview of our analogy based on the topological similarities of QCD and gravity.

\section{Neutrino Mass Generation}\label{sec:MassGen}

\begin{figure}[!t]
  \centering
      \includegraphics[width=8.6cm]{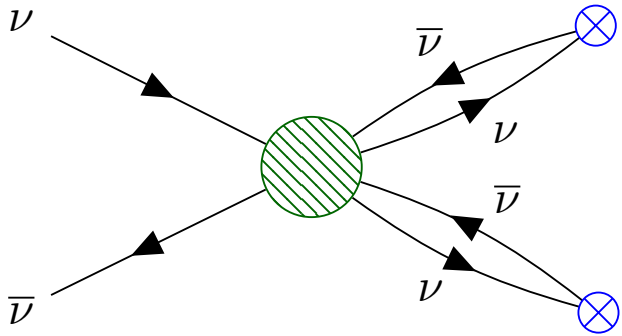}
  \caption{Neutrino mass generation through the condensate (crossed blue circles) via nonperturbative interaction (striped green circle).}
  \label{fig:MassGen}
\end{figure}

With the previous discussions of a neutrino vacuum condensate $\langle \nu \bar{\nu}\rangle\equiv v$ and the fluctuations around it, the pseudoscalar degrees of freedom $\phi$, we can write our composite neutrino field as
\begin{equation}
\nu \bar{\nu}=\langle \nu \bar{\nu}\rangle e^{i\phi}=v e^{i\phi}.
\end{equation}
We obtain an interaction between the $\phi=\{\phi_k,\eta_\nu\}$ and the neutrinos, 
\begin{equation}\label{eq:PhiCoupling}
\Lag \supset g_\phi \sum_k (\partial_\mu \phi_k \bar{\nu} \gamma^\mu\gamma_5 \nu) + g_{\eta_\nu} \eta_\nu \bar{\nu} \gamma_5 \nu,
\end{equation}
and an effective neutrino mass term 
\begin{equation}
\Lag \supset g_v v \bar{\nu}\nu.
\end{equation}
There are three important points to notice:
\begin{enumerate}[label={(\arabic*)}]
\item The mass term is not forbidden by any symmetries since chiral symmetry is broken by the chiral neutrino condensate formed through nonperturbative gravitational effects, as shown in Fig. \ref{fig:MassGen}. 
\item We are not making any extra assumption when pointing out that the neutrino condensate provides effective neutrino masses. This anomalous neutrino mass generation is exactly equivalent to the quark mass generation from the QCD condensate: the 't Hooft determinant $e^{i\theta}\det(\bar{q}_L q_R)+c.c.$ \citep{'tHooft1986} provides an effective anomalous coupling due to integrated instantons, which can be written as an effective four-fermion interaction.
\item By now we have treated the neutrinos as Dirac particles. If no right-handed neutrinos are introduced in the Standard Model (SM), the whole analysis remains similar, except the neutrino condensate is formed in the Majorana channel, $\langle \nu_L c\nu_L\rangle$, which breaks the lepton number spontaneously. Correspondingly, the resulting neutrino masses are of Majorana type.
\end{enumerate}

Since the gravitational topological vacuum susceptibility \eqref{eq:NuVacuum} only depends on the hard neutrino masses and not on our effectively generated masses, the correlator gets screened by $m_\nu=0$. Consequently, our model predicts that we will not observe any \textit{CP} violation in the gravitational vacuum\footnote{Remember that in QCD we can treat the quark masses as hard masses since the electroweak symmetry breaking occurs on a higher scale than the chiral symmetry breaking of QCD. This is not the case in our modified neutrino sector, where the scale $\Lambda_G$ of chiral symmetry breaking coincides with the scale of the vacuum condensate generating the masses.}. By analogy, it has been considered that the strong \textit{CP} problem could be resolved by assuming that the $u$ quark has a vanishing bare mass and only obtains a mass through the 't Hooft vertex \citep{Banks1994}\footnote{However, chiral perturbation theory indicates the need for an additional contribution to the up-quark mass and thus this solution to the strong $CP$ problem is apparently ruled out (for a discussion of this issue see \cite{Brambilla2014}).}.

In the following we will denote $m_\nu$ as the \textit{effective} neutrino mass rather than the hard one. In order to naturally provide an effective neutrino mass at the observed low-energy scale \citep{Ahmad2002}, the neutrino condensate needs to have a vacuum expectation value $v$ of order of the neutrino masses. The symmetry breaking scale $\Lambda_G$ coincides with $v$ as in the QCD analog, and hence also the $\eta_\nu$ mass has to be of the same order,
\begin{equation}
\Lambda_G\sim v\sim m_\nu \sim m_{\eta_\nu},
\end{equation}
where the largest neutrino mass can only be larger than the neutrino condensate by a maximal factor of $g_v= 4\pi$. 

It is interesting to notice that our neutrino vacuum condensate provides a vacuum energy at the scale of dark energy (the numerical coincidence of the neutrino mass scale and the dark energy scale has been pointed out before, see, e.g., \citep{Moffat2003,Bhatt2009,Azam2010}). On the first sight, our condensate seems to be not significantly different from other possible but not realized SM vacuum energy contributions; however, it is the only one triggered by nonperturbative gravitational effects.

If we finally include all three neutrino flavors, as already broached in the previous subsection, one immediate issue that we face in trying to generate all the neutrino masses effectively is how to generate the hierarchy of masses. In analogy with QCD, it may be expected that in the limit of zero bare neutrino masses the condensate should be universal in all the flavors and break $U(3)_L \times U(3)_R$ chiral symmetry to a diagonal $U(3)_{V}$ subgroup. That means the condensate would be a unit matrix in flavor space, $\langle \bar{\nu}_L \nu_R \rangle = \Lambda_G^3 \;{\rm diag} (1,1,1)$. 

We would like to stress that this is a detailed dynamical question, and a different pattern is equally possible. We can parameterize the patterns of symmetry breaking in very general terms by denoting the neutrino condensate  order parameter as a matrix in flavor space, $\langle \bar{\nu}_{\alpha_L} \nu_{\alpha_R} \rangle \equiv\hat{X}_{\alpha_L}^{\alpha_R}$, where $\alpha_L = 1,2,3$ and $\alpha_R = 1,2,3$ stand for the left- and right-handed flavor index, respectively. The effective potential for $\hat{X}$ is then some generic function of all possible invariants, for example,
\begin{equation}
V(\hat{X}) =  \sum_n {1 \over {n}} c_{2n} \Tr[(\hat{X}^+\hat{X})^n],
\end{equation}
where $c_{2n}$ are some coefficients. For simplicity we have excluded other invariants. The matrix $\hat{X}$ can always be brought to a diagonal form by a $U(3)_L \times U(3)_R$ rotation, $\hat{X} = {\rm diag}(x_1,x_2,x_3)$. The extremum values are then determined by the following set of equations:
\begin{equation} 
  {\partial V \over \partial x_j} =   x^*_j \left( \sum_n c_{2n} |x|_j^{2(n-1)} \right) = 0.
\label{roots} 
\end{equation}
It is clear that the vacuum expectation values are determined as the roots of the polynomial in the brackets and can be different and even hierarchical depending on the parameters $c_{2n}$. The symmetry breaking $U(3)_L \times U(3)_R \rightarrow U(3)_{V}$ corresponds to a particular choice of a single root, $x_1=x_2=x_3$. While this choice is conventionally assumed to be realized in QCD, there is no {\it a priori} reason to expect the same in other cases.

\section{\label{sec:PhenImplications}Phenomenological implications}

As we presented in the Introduction, numerous hidden beyond Standard Model (BSM) interactions and (pseudo)scalar degrees of freedom in the neutrino sector have been investigated to date. There are several constraints on the consequences of these models, which have been frequently updated in the past and will be further enhanced by future experiments. 

To present one example, the SM predicts an effective number of neutrino species in the early Universe of $N_{\rm eff} = 3.046$ \citep{Mangano2005}. While earlier observational values (e.g. $N_{\rm eff} = 3.14^{+0.70}_{-0.65}$ \citep{Cyburt2004} or $N_{\rm eff} = 4.34^{+0.86}_{-0.88}$ \citep{Komatsu2011}) still opened the window for many BSM predictions for $N_{\rm eff}$, recent Planck data ($N_{\rm eff} = 3.15\pm 0.23$ \citep{Ade2015}) narrows the window around the SM value and excludes many of the investigated scenarios. 

As we will discuss in the following first subsection, our symmetry breaking scale $\Lambda_G$ is only weakly constrained from the electroweak Higgs effect and from cosmological data. Therefore, we are allowed to predict numerous phenomenological consequences of our model, which shed new light onto important aspects of the history and content of our Universe.

In the subsequent subsections, we will describe these phenomenological aspects in detail. We will first point out the consequence of our neutrino mass generation mechanism for the cosmological neutrino mass bound in Sec. \ref{sec:MassBound}. In the subsequent Sec. \ref{sec:Bremsstrahlung}, we will evaluate low-energy $\phi$ radiation arising in high-energy neutrino processes and possibly from decays of disoriented chiral condensates. Section \ref{sec:Decays} consists of the enhancement of neutrino decays, and Sec. \ref{sec:Interaction} covers the fate of the cosmological neutrino background as well as neutrino cold dark matter in the late Universe. Finally, in Sec. \ref{sec:KATRIN}, we illustrate the possible observational consequences of a neutrino condensate and neutrino-neutrino interactions in terrestrial experiments.

\subsection{Bounds on symmetry breaking scale}\label{sec:BoundOnScale}

A universal upper bound on the symmetry breaking scale $\Lambda_G$ comes from the fact that the gravitationally triggered chiral fermion condensate contributes into the SM Higgs condensate. Such a chiral fermion condensate must also involve other fermion flavors besides the light neutrinos, including quarks and charged leptons. Indeed, since the leading-order gravity effects should distinguish the different fermions of the SM only by their masses, we predict that all the fermion flavors condense. Then the condensates of the light flavors $f$ of mass $m_f \ll \Lambda_G$ must be of order $\Lambda_G$, whereas the condensates of the heavy flavors $f$ of mass $m_f \gg \Lambda_G$ scale as $\langle \bar{f}_L f_R\rangle \sim \Lambda_G^4 / m_f$. Here we assume that heavy flavors decouple in the same way as in QCD, which may not be true in gravity. 

The immediate bound on $\Lambda_G$ comes from the fact that such a condensate, similarly to the quark condensate in QCD, contributes into the Higgs effect of the $W$ and $Z$ bosons. This immediately tells us that $\Lambda_G$ must be below the QCD scale. For $\Lambda_G \sim m_{\nu}$, the contribution to the SM Higgs effect is negligible and the small contamination of the Higgs condensate by the condensate of the quarks and leptons is in accordance with electroweak constraints. However, it would be interesting to see if such a gravity-induced condensate, say, for charged leptons, $\langle \bar{e}_L e_R\rangle = \Lambda_G^4/m_e$, can have some potentially observable effects that may be probed experimentally.

Further constraints on the symmetry breaking scale and on the BSM modifications of the neutrino sector come from cosmology. The most important cosmological restriction is the neutrino free-streaming constraint at the photon decoupling epoch, where the temperature of the Universe was $T\sim 256~{\rm meV}$ \citep{Hinshaw2009}. The authors of \citep{Sellentin2014} found that neutrino free-streaming is favored by the data over a relativistic perfect neutrino fluid with $\Delta \chi^2 \simeq 21$. Also secret Yukawa couplings between neutrinos and BSM low-mass Majoron-like (pseudo)scalars such as our $\phi$'s are strongly restricted in the early Universe \citep{Hannestad2005,Basboll2008,Archidiacono2013}. For example, if the neutrinos already coupled to the $\phi$'s before $T\sim 256~{\rm meV}$, their diagonal interactions would have to be smaller than $1.2\times 10^{-7}$, and their off-diagonal couplings would be even stronger constrained in order to be consistent with the observations \citep{Archidiacono2013}. 

Due to the small neutrino mass splitting \citep{Olive2014} we assume $\Lambda_G$ and thus the temperature of the phase transition to be below $256~{\rm meV}$, leaving our model cosmologically unconstrained. \textit{A priori}, $\Lambda_G$ could also be larger than $256~{\rm meV}$. However, even if we ignore the cosmological neutrino mass limit and consider neutrino masses of up to 2.2~eV \citep{Wolf2010}, as outlined in Sec. \ref{sec:MassBound}, generating these masses through couplings to a condensate of scale $\Lambda_G< 256~{\rm meV}$ would be perfectly consistent. This is because the diagonal couplings of the neutrinos to the condensate are observationally unconstrained and only the off-diagonal couplings are experimentally restricted, leaving open all possible couplings $g_v\leq 4\pi$, as we will point out in Sec. \ref{sec:Decays}. Therefore, we assume a phase transition in the very late Universe, $\Lambda_G< 256~{\rm meV}$, leading to no cosmological restrictions on our modifications of the neutrino sector.

The smallest possible mass of the heaviest neutrino, $m_{\nu_{\rm heavy}}\sim 50~{\rm meV}$ \citep{Olive2014}, and the maximal coupling of the condensate to this neutrino, $g_v= 4\pi$, requires at least a scale of $\Lambda_G\sim 4 ~{\rm meV}$. In the following we will thus denote the scale of the phase transition as $\Lambda_G\sim$~meV, which will serve as a short-hand notation for a possible range of $\Lambda_G$ between $4$ and $256~{\rm meV}$.

\subsection{Invalidity of cosmological neutrino mass bound}\label{sec:MassBound}

The recent idea of a neutrinoless universe \citep{Beacom2004} sparked a lot of excitement among cosmologists since it eludes the cosmological neutrino mass bound emerging from large scale structure. However, the model was finally ruled out by the aforementioned precision measurements of the effective number of neutrino species in the early Universe \citep{Ade2015}. 

Our model predicts that the neutrinos had been massless until the phase transition in the very late Universe. For $v=\Lambda_G=T_{\Lambda_G}$, the temperature of the phase transition is in the range of $4~{\rm meV} \lesssim T_{\Lambda_G} \lesssim256~{\rm meV}$, corresponding to a redshift of $16 \lesssim z \lesssim 1089$. Due to enhanced neutrino decays (see Sec. \ref{sec:Decays}), directly after the phase transition, all relic neutrinos decay into the lightest neutrino mass eigenstate. 

Thus, our model gives rise to the same phenomenological consequence, the vanishing cosmological mass limits, $\sum_\nu m_\nu < 0.28~{\rm eV}$ \citep{Thomas2010} or even $\sum_\nu m_\nu < 0.18~{\rm eV}$ \citep{Riemer-Sorensen2013}. Depending on the exact time of the phase transition, either no information on the neutrino masses can at all be inferred from cosmology, or the cosmological mass bound only applies to the lightest neutrino species, which in principle can even be massless\footnote{Note that the authors of \citep{Battye2013} claim to have cosmologically detected nonzero neutrinos masses $\sum_\nu m_\nu = (0.320\pm 0.081)~{\rm eV}$ for the degenerate mass scenario with $4\sigma$.}.

Notice here that in the late Universe, the neutrinos are not in thermal equilibrium anymore, and thus one might expect that the phase transition and the mass generation may not happen as in the thermal case. However, the neutrino energy density should affect the order parameter in a way similar to the temperature. As long as the energy density scales as $T^4$, the neutrinos will obtain their mass at the scale $\Lambda_G\sim {\rm meV}$. 

According to our model, only terrestrial experiments are currently suitable to determine the absolute neutrino mass scale. The latest of these experiments still allows neutrino masses of up to $2.2~{\rm eV}$ \citep{Wolf2010}. If our model appears to be valid and the cosmological constraints do not hold true, the KATRIN experiment could in principle discover the neutrino masses down to $0.2~{\rm eV}$ \citep{Drexlin2013} in the near future.

\subsection{Creation of new particles in high-energy neutrino processes}\label{sec:Bremsstrahlung}

Since the masses of the $\phi$'s are either zero for the Goldstones or are determined by the low-energy scale $\Lambda_G\sim {\rm meV}$ for $\eta_\nu$, these new particles could in principle be created in every high-energy neutrino process. Analogous to pion bremsstrahlung \citep{Weinberg1970}, the $\phi$ radiation spectrum would be continuous and would have a peak at small energies above $m_{\phi}$. This process would be much more important than the usual highly suppressed bremsstrahlung emittance of Z bosons.

By analogy with the disoriented chiral condensate (DCC) in QCD \citep{Mohanty2005}, we might also have such condensates in high-energy neutrino processes. The DCC may finally decay into the real vacuum by emission of coherent low-energy $\phi$'s.

All these production processes of $\phi$ particles could have a strong impact on high-energy neutrino processes, most importantly on star cooling. Supernova cooling restricts the diagonal coupling of Majorons to neutrinos \citep{Fuller1988,Berezhiani1989,Kachelriess2000,Farzan2002}, but since our interactions are not necessarily lepton number violating and introducing the $\phi$'s does not open up a new emission channel, none of the current constraints hold true for our couplings of the $\phi$'s to neutrinos.

If $\eta_\nu$ is heavier than the lightest neutrino, it would immediately decay and would not be observable. The hidden numerical parameter of the anomaly in Eq. \eqref{eq:etamassLag} and the exact scale $\Lambda_G$ are unknown, implying that the absolute mass of $\eta_\nu$ cannot be predicted. If the mass $m_{\eta_\nu}$ is lower than the lightest neutrino, this new degree of freedom could probably be detected in future experiments, as well as the other $\phi_k$ Goldstones.

\subsection{Enhanced neutrino decays}\label{sec:Decays}

Conventional neutrino interactions only provide decays of heavier neutrinos into lighter ones, which are suppressed by the $W$ and $Z$ boson masses, leading to neutrino lifetimes which exceed the lifetime of the Universe. 

As already pointed out in Sec. \ref{sec:MassGen}, the massive $\eta_\nu$ and the massless Goldstone particles $\phi_k$ open up new decay channels for the neutrinos,
\begin{equation}
\hspace*{-0.1cm} \Lag \supset  \sum_k \partial_\mu \phi_k \sum_{ij} g_{\phi,ij} \bar{\nu}_i \gamma^\mu\gamma_5 \nu_j + \eta_\nu \sum_{ij} g_{\eta_\nu,ij} \bar{\nu}_i \gamma_5 \nu_j.
\end{equation}
Here it is important to notice that for the neutrino decays $\nu_i\rightarrow\nu_j+\phi$ and $\nu_i\rightarrow\bar{\nu_j}+\phi$ (where $m_i>m_j$), the pseudoscalar and derivative couplings are equivalent \citep{Hannestad2005}. Therefore, we will for simplicity assume only pseudoscalar couplings and will denote all couplings between the $\phi$'s and the neutrinos as $g_{ij}$ in the following.

Our enhanced neutrino decays can happen via intermediate $\phi$ states, i.e., via box diagrams or Fermi-like interactions. However, the resulting decay widths are suppressed by four powers of the off-diagonal couplings, $g_{ij}^4$, just like the neutrino annihilation, which we will discuss in the next subsection.

The process, in which a physical $\phi$ particle is emitted, is only suppressed by $g_{ij}^2$. The decay rate $\Gamma_i$ of the sum of the two processes $\nu_i\rightarrow\nu_j+\phi$ and $\nu_i\rightarrow\bar{\nu}_j+\phi$ in the rest frame of $\nu_i$ is \citep{Beacom2002,Hannestad2005}
\begin{equation}\label{eq:NDecay}
\Gamma_i=g_{ij}^2m_{i}
\end{equation}
if we neglect the masses of the final states and omit numerical factors of order $\mathcal{O}(10^{-2})$. In the medium frame the rate is reduced by a Lorentz factor of $m_i/E$.

This transfers for the lowest possible normal-ordered masses of $m_1=0~{\rm meV}$, $m_2=9~{\rm meV}$, and $m_3=50~{\rm meV}$ \citep{Olive2014} into the neutrino rest-frame lifetimes $\tau_i=1/\Gamma_i$ of
\begin{align}
\frac{\tau_3}{m_3}&\simeq \frac{1\times 10^{-11}}{g_{32+31}^2}\frac{{\rm s}}{{\rm eV}},\\
\frac{\tau_2}{m_2}&\simeq \frac{4\times 10^{-10}}{g_{21}^2}\frac{{\rm s}}{{\rm eV}}.
\end{align}
As already mentioned, for $\Lambda_G>256~{\rm meV}$, such a modification by secret Majoron-type interactions would be highly constrained by CMB data \citep{Hannestad2005,Archidiacono2013}, but for our symmetry breaking scale of $\Lambda_G\sim {\rm meV}$, these cosmological constraints do not hold true. However, constraints from accelerator, atmospheric, and solar neutrino experiments play an important role since our enhanced decays take place on all energy scales. 

The current \citep{Gomes2015} noncosmological experimental constraints on the neutrino mass eigenstate lifetimes for a normal nondegenerate mass hierarchy are \citep{GonzalezGarcia2008,Eguchi2004}
\begin{align}
\frac{\tau_3}{m_3}&>9\times 10^{-11}\frac{{\rm s}}{{\rm eV}},\\
\frac{\tau_2}{m_2}&>1\times 10^{-3}\frac{{\rm s}}{{\rm eV}},
\end{align}
which enforce our off-diagonal couplings to be 
\begin{align}\label{eq:CouplingLimits}
g_{32+31}\lesssim 4\times 10^{-1}\;\:\:\:\:\text{and}\;\:\:\:\:g_{21}\lesssim 6\times 10^{-4}.
\end{align}

Standard flavor oscillations imply an expected neutrino flavor ratio $(\nu_e:\nu_\mu:\nu_\tau)$ of $(1:1:1)$. Our enhanced decays of the heavier into the lightest neutrino would lead to the dominant presence of a distinct flavor composition in long-traveling extraterrestrial neutrino fluxes. As observed in \citep{Hannestad2005}, with an assumed neutrino flux of energy $E=10~{\rm TeV}$ coming from a source at distance $D=100~{\rm Mpc}$, a strong decay effect is visible if $\Gamma_i (m_i/E)\agt D^{-1}$. Taking into account Eq. \eqref{eq:NDecay}, this means that couplings of 
\begin{equation}\label{eq:gDecay}
g_{ij}\agt1\times 10^{-7}\left(\frac{50{\rm ~meV}}{m_i}\right) \left(\frac{E}{{\rm 10~TeV}}\right)^{1/2} \left(\frac{{\rm 100~Mpc}}{D}\right)^{1/2}
\end{equation}
already lead to observable effects. Our constraints on the couplings \eqref{eq:CouplingLimits} therefore imply that the expected deviation from an equal neutrino flavor content could be measured in extraterrestrial neutrino fluxes detected, for example, with the IceCube experiment. 

With the three-year data of the IceCube experiment, an equal flavor composition is excluded at 92\% C.L. by one analysis \citep{Mena2014}, and the best fit is obtained for a ratio $(\nu_e:\nu_\mu:\nu_\tau)$ of $(1:0:0)$. This matches our idea that the heavier neutrinos decay into the lightest neutrino, which would be mainly composed out of $\nu_e$ neutrinos in case of a normal mass hierarchy \citep{King2013}. Another analysis gives the best-fit ratio of $(0:0.2:0.8)$ but also an equal flavor ratio or a ratio of $(1:0:0)$ are not significantly excluded \citep{Aartsen2015}. A dominance of $\nu_\mu$ and $\nu_\tau$ neutrinos over $\nu_e$ neutrinos would match our decay picture in case of an inverted mass hierarchy \citep{King2013}.

In order to obtain significant results for a deviation from an equal flavor ratio, more data is needed. If our model is true, equal flavor ratios are ruled out since the state $\nu_2$ with nearly equal flavor content cannot be the lightest mass state \citep{Shrock2002}. Therefore, our predicted enhanced neutrino decays can probably be verified in the near future.

If enhanced neutrino decays will be observed, this modification of neutrino physics will also play an important role in modeling supernova (SN) events. Most crucially, the enhanced decays would imply that the neutrinos from SN 1987A \citep{Hirata1987,Bionta1987} have decayed into the lightest mass eigenstate on their way to Earth. Since SN 1987A was about $D=50~{\rm kpc}$ away from Earth \citep{Panagia1999} and the neutrino flux energy was in the range of $E=10~{\rm MeV}$ \citep{Mirizzi2015}, decay effects would have already occurred for off-diagonal couplings of \eqref{eq:gDecay}
\begin{equation}\label{eq:gDecaySN}
g_{32+31}\agt 4\times 10^{-9}\;\:\:\:\:\text{and}\;\:\:\:\:g_{21}\agt 2\times 10^{-8},
\end{equation}
where we again assume the lowest possible normal-ordered neutrino mass scheme. 

If our proposed neutrino decays are mediated by off-diagonal couplings in the range given by Eqs. \eqref{eq:CouplingLimits} and \eqref{eq:gDecaySN}, the analyses of the original neutrino spectra of SN 1987A and specifically the constraints on the flavor composition of the observed neutrinos \citep{Lunardini2004} have to be substantially modified.\footnote{The authors thank one of the anonymous referees for this remark.} This decay scenario is not excluded so far, since the SN 1987A data restricts only the lowest mass eigenstate to be stable, $\tau_1/m_1>10^5~{\rm s/eV}$ \citep{Baerwald2012}, and the simulations of supernova explosions still exhibit many uncertainties \citep{Arnett1989,Lindner2001,Ando2003,Ando2004,Fogli2004}.

\subsection{Fate of cosmic neutrino background and neutrino cold dark matter}\label{sec:Interaction}

The neutrino vacuum condensate may require free neutrinos below meV energies to bind up, as discussed in Sec. \ref{sec:GravityMassGap}. Consequently, the relic neutrinos in our current Universe would all be in massless Goldstone bound states. However, in the following we want to consider other possible fates of the cosmological neutrino background in case this proposal does not hold true. 
 
Relic neutrinos and antineutrinos can annihilate into massless Goldstones with an annihilation rate in the non-relativistic limit of \citep{Beacom2004}
\begin{equation}\label{eq:Gamma}
\Gamma(T)=\langle \sigma v\rangle n_{\rm eq}=g^4T\left(\frac{T}{m_\nu}\right)^{3/2}e^{-m_\nu /T},
\end{equation}
where $\sigma$ is the annihilation cross section, $v$ is the neutrino velocity, $n_{\rm eq}$ is the neutrino equilibrium density, and $g$ is the diagonal or off-diagonal coupling. We omit numerical factors of order $\mathcal{O}(10^{-4})$. 

The enhanced neutrino decays imply that all the relic neutrinos immediately after the phase transition decay into the lightest mass eigenstate $\nu_1$, where we again suppose normal mass hierarchy. In the following we will present two scenarios which make clear that the explicit scales of $m_1$ and $\Lambda_G$ are crucial for the subsequent fate of the cosmological neutrino background.

Let us in the first scenario assume a quite heavy lowest-mass eigenstate, $m_1\sim 50~{\rm meV}$, and a phase transition at $T_{\Lambda_G}=\Lambda_G\gtrsim 50~{\rm meV}$. As argued in \citep{Beacom2004}, our enhanced interactions after $T_{\Lambda_G}$ keep the neutrinos in equilibrium until $T_\nu<m_1$. Afterwards the neutrino abundance will undergo exponential suppression until the annihilation rate $\Gamma(T)$ \eqref{eq:Gamma} becomes equal to the Hubble expansion rate $H(T)$, i.e., the neutrinos freeze out. If the freeze-out temperature is $T_f<\mathcal{O}(m_1/7)$, the neutrino abundance becomes negligible due to exponential suppression by a factor $>100$. Solving $\Gamma(T_f)\equiv H(T_f)$ on condition that $T_f<m_1 /7$ hence provides a constraint on the minimal coupling $g$ which is necessary for obtaining a ``neutrinoless universe''. We find that a coupling of $g\gtrsim 3\times 10^{-6}$ is required in order to annihilate a significant amount of neutrinos into Goldstone bosons by $T_f$, leaving only a negligible fraction of relic neutrinos behind.

If we consider a second scenario in which the mass of the lightest neutrino is negligibly small, $m_1\lesssim 1~{\rm meV}$, we find that for a coupling of $g\gtrsim 2\times 10^{-6}$, the neutrino annihilation rate would still be higher than the expansion rate today, $\Gamma(T_{0})>H(T_{0})$. In this case, the relic neutrinos of the late Universe have not frozen out until today. These strongly coupled low-energy neutrinos then either behave as a superfluid or form bound states, where both options are perfectly compatible with all the current observations.  Then the phase transition at $T_{\Lambda_G}=\Lambda_G$ implies a transition of neutrino hot dark matter to neutrino cold dark matter. This cold dark matter, made up of bound neutrino states or a neutrino superfluid, could be partially responsible for the galaxy rotation curves in our Universe. An additional contribution to cold dark matter could be due to $\eta_\nu$ particles and topological defects, such as textures (neutrino Skyrmions) or global monopoles, which may have formed during neutrino condensation. However, still additional cold dark matter is required in order to explain, \textit{inter alia}, earlier structure formation and CMB data.

\subsection{Signals in terrestrial experiments}\label{sec:KATRIN}

One terrestrial neutrino experiment, which is of importance for our model, is the KATRIN experiment starting to operate this year. In this experiment, the kinetic energy spectrum of an electron created in tritium beta decay is measured. The end point of this spectrum depends on the neutrino mass, parametrized in terms of the phase space factor, $\sqrt{(E_0-E)^2-m_{\nu_e}^2}$, where $E_0$ is the maximal electron energy (the end of the spectrum if we had no neutrino mass), and $E$ is the real variable kinetic energy of the electron \citep{Drexlin2013}. This phase space factor could be altered by our modification of the low-energy neutrino sector.\footnote{The authors thank G. Raffelt for this suggestion.}

At this point it is important to mention that assuming neutrinos to be in bound states below the low-energy symmetry breaking scale $\Lambda_G\sim {\rm meV}$ would not constrain processes with single low-energy neutrino emission, since the singly emitted neutrinos would directly ``hadronize" in form of massless $\phi_k$-Goldstones by picking up partners from the neutrino sea. This may happen with the emitted antineutrinos at the end point of the beta decay spectrum at KATRIN, where the neutrinos have energies below $\Lambda_G$. 

The abundant cosmological neutrinos may be detectable through neutrino capture on the radioactive nuclei, so that we get a disctinct monoenergetic electron energy peak above the initial end $E_0$ of the electron energy spectrum, $E=E_0+m_{\nu_e}$. This idea was first elaborated in \citep{Weinberg1962} and further evaluated in \citep{Kaboth2010} and \citep{Cocco2007}, among others. 

This distinct signal should in principle be detectable and seperable from the end of the continuous decay spectrum, but the detector sensitivity to measure this deviation is not high enough for the current low density of relic neutrinos in the Universe. Gravitational clustering of the neutrinos in our Galaxy could lead to a maximal local neutrino overdensity of $10^6$, which would result only in $1.7$ counts by KATRIN per year, and hence would not be detectable \citep{Faessler2015}. However, our new neutrino-neutrino interactions as well as the abundant $\phi$'s and neutrino Skyrmions, which have possibly formed during the phase transition, could lead to an enhanced neutrino overdensity of more than $2\times 10^9$, which is the lower limit for a detection at KATRIN \citep{Kaboth2010}.

Finally, we would like to note that if axial lepton number is weakly gauged, e.g., in form  of $B-L$ local symmetry, the neutrino condensate would trigger a mass for the $B-L$ gauge boson. This could result in some interesting experimental prospects by looking for signatures of the $B-L$ force in short-distance measurements, such as the ones presented in \citep{Adelberger2009}. The existence of a gravity-competing force in form of a gauged $B-L$ symmetry was originally suggested in the context of large extra dimensions \citep{ArkaniHamed1998_2}. In the present context however the allowed parameter range is different. It would be interesting to explore this question in future.

\section{Discussion and conclusions}\label{sec:Conclusion}

In the current paper, we presented how neutrino condensation and small neutrino masses directly emerge from a topological formulation of gravitational anomaly. In order to clarify our argument, we first outlined the analogy to well-known QCD effects. Based on \citep{Dvali2005_2, Dvali2005, Dvali2013}, we recapitulated that gravity and QCD have a very similar topological and anomaly structure.  This similarity relies on only one assumption: that the gravitational $\theta$-term is physical. In the presence of chiral fermions, the elimination of the vacuum $\theta$-angle through chiral anomaly then amounts to the generation of a mass gap. Consequently there exists a bound neutrino state $\eta_\nu$ triggered by gravitational anomaly, analogous to the $\eta'$ triggered by the ABJ anomaly of QCD. 

As we showed, this predicted new bound neutrino state implies important theoretical consequences for the neutrino sector: a neutrino vacuum condensate has to emerge for consistency reasons. Without making any additional assumptions, we pointed out that this vacuum condensate generates the observed small neutrino masses, through interactions mediated by the same nonperturbative gravitational effects which are responsible for the gravitational anomaly. 

We reminded the reader that the connection between the topological vacuum susceptibility and a condensate in QCD was proven to be unrelated to confinement. Moreover, we found that the QCD computations of the topological vacuum susceptibility are in full accordance with our derivation in the gravity sector. We also pointed out that neutrinos below energy scales $\Lambda_G$ may have to be in bound states.

We emphasized that our mass generation mechanism can satisfy the observed neutrino mass hierarchy and is independent of the Majorana or Dirac nature of the neutrinos. Furthermore, we explained that in case where all neutrino masses are exclusively generated by our effective mechanism, not only does one new degree of freedom $\eta_\nu$ emerge, but up to 14 massless Goldstone bosons $\phi_k$ emerge as well, analogous to the pseudoscalar mesons in QCD.

In the penultimate section, we investigated the phenomenological consequences of our model:

\begin{enumerate}[label={(\roman{enumi})}]
\item Not only the neutrinos but also all other fermion flavors condense, which leads to constraints on the symmetry breaking scale $\Lambda_G$ from the electroweak Higgs effect. Due to the experimental lower bound on neutrino masses and cosmological constraints on BSM neutrino physics, the scale $\Lambda_G$ has to be in the range of $4~{\rm meV} \lesssim \Lambda_G \lesssim 256~{\rm meV}$.
\item The cosmological neutrino mass bound vanishes since the neutrinos had been massless until the very late Universe, $T=\Lambda_G$. Therefore, neutrino masses of up to $2.2~{\rm eV}$ are still allowed, and the KATRIN experiment may detect them in the near future.
\item Coherent radiation of $\phi$ particles, i.e., massive $\eta_\nu$'s as well as massless Goldstones $\phi_k$, is emitted in high-energy neutrino processes and possibly in decays of disoriented chiral condensates. This radiation can be detected in future experiments.
\item After the phase transition at $T=\Lambda_G$, all relic neutrinos directly decay into the lightest mass eigenstate. Deviations from an equal flavor rate due to neutrino decays in extraterrestrial neutrino fluxes can be observed in future IceCube data. The enhanced neutrino decays may also necessitate modified analyses of the original neutrino spectra of the supernova SN 1987A.
\item Depending on the exact scale $\Lambda_G$ and on the smallest neutrino mass, the predicted strong neutrino-neutrino interactions have two different consequences for the relic neutrinos: either they completely annihilate into massless Goldstone bosons, or they bind up or form a superfluid. In the latter scenario, the relic neutrinos provide a contribution to cold dark matter in the late Universe, together with the $\eta_\nu$'s and topological defects, which may have formed during neutrino condensation.
\item The enhanced neutrino-neutrino interactions could also lead to relic neutrino clustering in our Galaxy. The clustering as well as the abundant $\phi$'s and neutrino Skyrmions from the phase transition could lead to a neutrino overdensity of more than $2\times 10^9$, which would make the cosmic neutrino background detectable in the KATRIN experiment. 
\item If axial lepton number is $B-L$ gauged, the neutrino condensate would provide a mass for the $B-L$ gauge boson, leading to a gravity-competing force detectable in short-distance measurements.
\end{enumerate}

With the correlator $\langle R\tilde{R}\rangle$ \eqref{eq:NuVacuum} we have introduced a new scale $\Lambda_G$ into gravitational physics. The theoretical origin of this new scale induced by nonperturbative gravitational effects will be further investigated by the authors.

Concerning this scale, we observe a crucial difference of the gravitational scenario to the QCD analog: the gravitational scale $M_{P}$ where gravitational interactions become strong cannot be equal to the scale of symmetry breaking $\Lambda_G$, which we identify with the neutrino mass scale. This seems to be different in QCD in which the symmetry-breaking scale is not far from the scale where perturbative gluon-gluon interactions get strong. In sharp contrast, the coupling of gravitons of wavelength $\Lambda_G^{-1}$,  given by  $\alpha_G \sim \Lambda^2_G / M_{P}^2$, is minuscule. This naive difference however should not confuse the reader. First, our analogy was based not on perturbative portraits, but rather on striking similarities between the topological and anomaly structures of the two theories. Second, the scale $\Lambda_G$ has to be understood not as a scale of perturbative strong coupling, but as the exponentially small scale where collective nonperturbative phenomena become important. 

Finally, we would like to discuss our model's implications on $CP$ violation in gravity. As shown in \cite{Dvali2013}, the $\eta_\nu$ boson presented in our scenario plays the role of the axion for the gravitational analog of the QCD $\theta$-term, since the gravitational chiral anomaly generates the coupling $\eta_{\nu} R\tilde{R}$ in the Lagrangian. This coupling relaxes the gravitational $\theta$-term to zero, preventing the manifestation of $CP$ violation by the gravitational vacuum. This effect is fully analogous to the QCD axion scenario, in which the pseudoscalar axion suppresses strong $CP$ violation by the vacuum $\theta$-angle.
 
However, the promotion of the gravitational $\theta$-angle into a dynamical field $\eta_{\nu}$ provides us with potentially observable $CP$-violating effects in out-of-vacuum processes, such as gravitational waves (GW). For example, the backgrounds with time-dependent $\eta_{\nu}$ can be probed by GW detections as suggested in \citep{Jackiw2003} in the context of Chern-Simons modified General Relativity (for a review see \citep{Alexander2009}). Recently, the LIGO Collaboration accomplished the first direct GW measurement \citep{Abbott2016}. However, their further analysis \citep{TheLIGOScientific2016} of the observed GW signal of a binary black hole merger so far yielded no constraints on the Chern-Simons modification of General Relativity, since there is a lack of related theoretical predictions for merging black hole GW signals, making such studies not yet feasible.

\section*{Acknowledgements}

G.\ D.\ and L.\ F.\ would like to thank Georg Raffelt, Mikhail Shifman, C\'{e}sar G\'{o}mez, Nico Wintergerst, Tehseen Rug, Andr\'{e} Fran\c{c}a, and Daniel Flassig for discussions about theoretical and phenomenological aspects of the model. The work of G. D. was supported in part by Humboldt Foundation under Humboldt Professorship, ERC Advanced Grant 339169 ``Selfcompletion'', by TR 33 ``The Dark Universe'', and by the DFG cluster of excellence ``Origin and Structure of the Universe''. The work of L.\ F.\ was supported by the International Max Planck Research School on Elementary Particle Physics.

\bibliographystyle{apsrev4-1}
\interlinepenalty=10000
\bibliography{Bibliography}

\end{document}